\documentclass[aps,showpacs,pra,twocolumn,twoside,groupedaddress]{revtex4}
\usepackage{graphicx}
\usepackage{epsfig}
\usepackage{epsf}
\usepackage{amssymb}
\usepackage{epstopdf}
\usepackage{amsmath}
\usepackage{url}
\usepackage{color}

\begin{document}

\title{Magnetic Resonance Lithography with Nanometer Resolution}

\author{Fahad AlGhannam$^{1,2}$, Philip Hemmer$^{3}$\footnote{prhemmer@ece.tamu.edu}, Zeyang Liao$^{1}$\footnote{zeyangliao@physics.tamu.edu}, and M. Suhail Zubairy$^{1}$\footnote{zubairy@physics.tamu.edu}}

\affiliation{$^1$Institute for Quantum Science and Engineering (IQSE)  and Department of Physics and Astronomy, Texas A$\&$M University, College Station, TX 77843-4242, USA\\
$^2$ The National Center for Applied Physics, KACST, P.O.Box 6086, Riyadh 11442, Saudi Arabia \\
$^3$ Department of Electrical $\&$ Computer Engineering,  Texas A$\&$M University, College Station, TX 77843-4242, USA}

\begin{abstract}
We propose an approach for super-resolution optical lithography which is based on the inverse of magnetic resonance imaging (MRI). The technique uses atomic coherence in an ensemble of spin systems whose final state population can be optically detected.  In principle, our method is capable of producing arbitrary one and two dimensional high-resolution patterns with high contrast.
\end{abstract}

 \maketitle

\section{introduction}

Optical lithography is widely used to print circuit images onto substrates because of its inherent parallelism and the fact that its non-ionizing illumination avoids substrate damage. However, the resolution of traditional optical lithography is restricted by the Rayleigh diffraction limit \cite{Rayleigh, Abbe, Review, Hemmer_review}. Due to this limit, we have to use shorter wavelengths such as extreme ultraviolet or X-ray to print smaller patterns \cite{Gwyn,Heuberger}. However, insulating materials such as silicon dioxide, when exposed to photons with energy greater than the band gap, release free electrons which subsequently can cause adverse charging which can lead to damage \cite{DOWLING}. It is therefore interesting and useful to invent an optical lithography scheme that can overcome the diffraction limit.

In the past two decades, a number of methods have been proposed including multiphoton absorption \cite{Yablonovitch,Peer,Bentley}, quantum entanglement \cite{Boto,Shih,Agarwal}, dopperlon \cite{Hemmer,Sun,Ge}, and dark states \cite{Kiffner}. However, these methods either require higher order multiphoton absorption, higher order quantum entangled states or multi atomic levels to write patterns with higher order resolution. Thus improving resolution more than ten times the optical wavelength with these methods is  difficult in practice. In 2010, we proposed a novel scheme to overcome the diffraction limit in optical lithography by inducing Rabi oscillations between two energy levels of the medium \cite{Liao1,Liao2,Liao3}. The extension from lower resolution to higher resolution is straightforward where we just need to increase the power of the excitation laser or the interaction time. However, an effective way to print arbitrary one dimensional(1D) and two dimensional (2D) patterns with low background is difficult in this method. Donut beam lithography \cite{Brimhall,Gan},which is the inverse of the ground state depletion super-resolution (GSD) technique \cite{Hell1,Hell2}, can write arbitrary patterns with resolution beyond the diffraction limit, but it requires point-by-point scanning instead of full frame exposure and is very inefficient since most of the laser light is actually used to suppress resist exposure. Using frequency-encoded light masks in neutral atom lithography it is possible to print arbitrary sub-diffraction-limited features \cite{Thywissen}. However, in atom lithography very high atom collimation is required for high resolution which is difficult to achieve with sufficient throughput to be useful for most lithography applications..     

In this paper, we propose an approach based on the MRI super-resolution technique \cite{Haacke}. Our method can be conveniently called magnetic resonance lithography (MRL) \cite{Hemmer_spie}. This method can be highly parallel, where the parallelism results from the fact that coherent excitation of two-level spin-flip transitions is employed. In this method, the extension from lower resolution to higher resolution is also straightforward where we just need to apply larger magnetic gradient and narrower bandwidth of microwave. Furthermore, it is also possible to print arbitrary 1D and 2D patterns with low background.

This paper is organized as follows. In Sec. II, we discuss the basic physics of MRL and propose the pulse sequence to print arbitrary 1D and 2D patterns. In Sec. III, we numerically demonstrate how to print an arbitrary one dimensional and two dimensional patterns into the spin states and the photoresist. In Sec. IV, we discuss some possible materials for MRL. Finally, we summarize the results.

\section{Basic principle and proposed experiment}

MRL requires a paramagnetic system with high spin transition quality factors and high contrast optical readout of the spin state. In addition, to write an arbitrary 2D patterns the system should have three magnetic sublevels. The proposed schematic setup is shown in Fig. 1(a) where the paramagnetic system is imbeded near the surface of the substrate. We can create the magnetic gradient fields using conductive metal wires fabricated on the substrate surface by means of a conventional photolithography process. A static magnetic field $B_{0}$ is applied from the perpendicular direction to uniformly split the magnetic sublevels $m=0$ and $m=\pm 1$ shown in Fig. 1(b). A microwave pulse with special spectral components is also applied from the perpendicular direction to excite the spin transitions $m=0\rightleftharpoons m=\pm 1$. After preparing the required pattern in the spin states, we may use a laser to excite the spins in state $m=0$ into a fluorescing excited state. A layer of photoresist covers the substrate's surface to capture the energy from the fluorescing excited state and the spin pattern can then be transfered to the photoresist. 

\begin{figure}
\includegraphics[width=0.58\columnwidth, height=3cm]{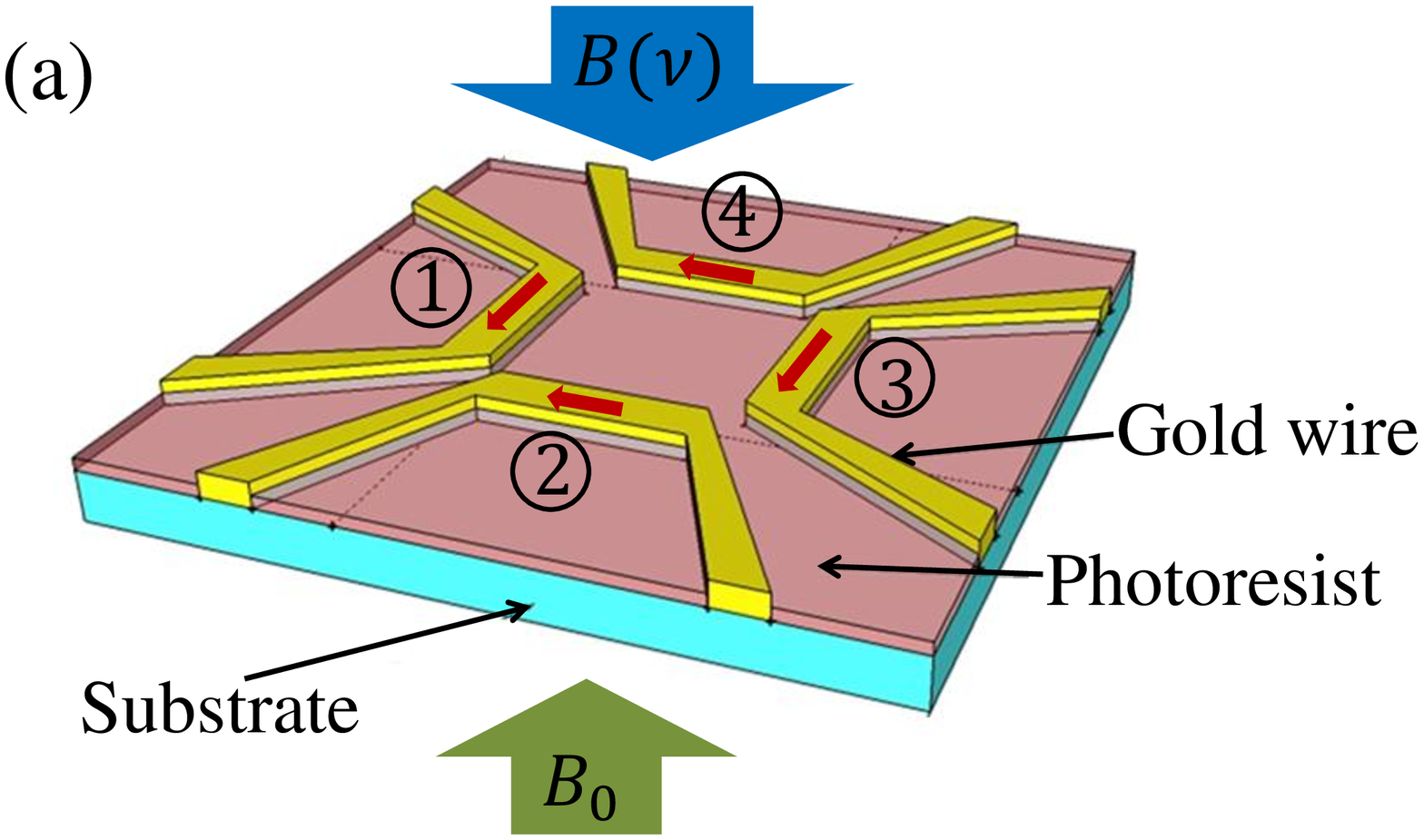}
\includegraphics[width=0.40\columnwidth, height=3cm]{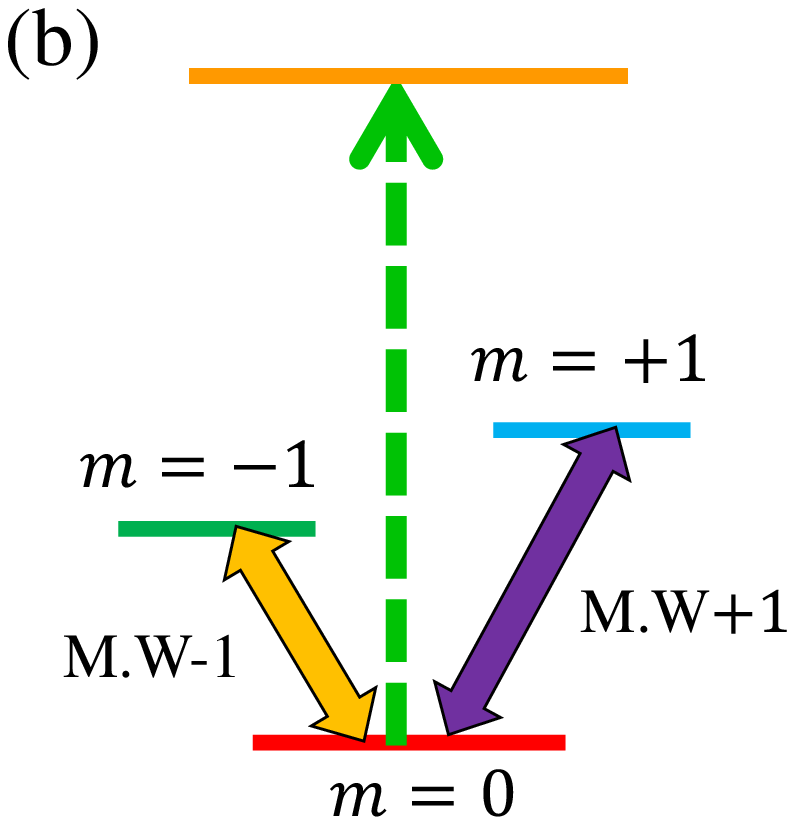}
\includegraphics[width=0.96\columnwidth]{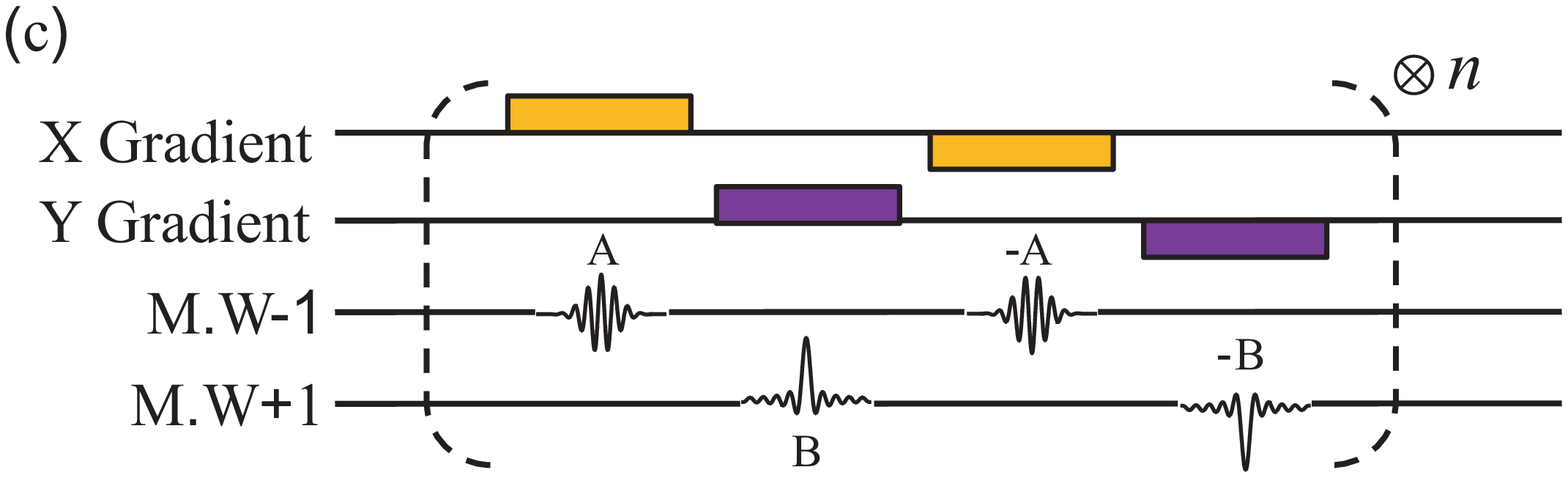}
\includegraphics[width=0.98\columnwidth]{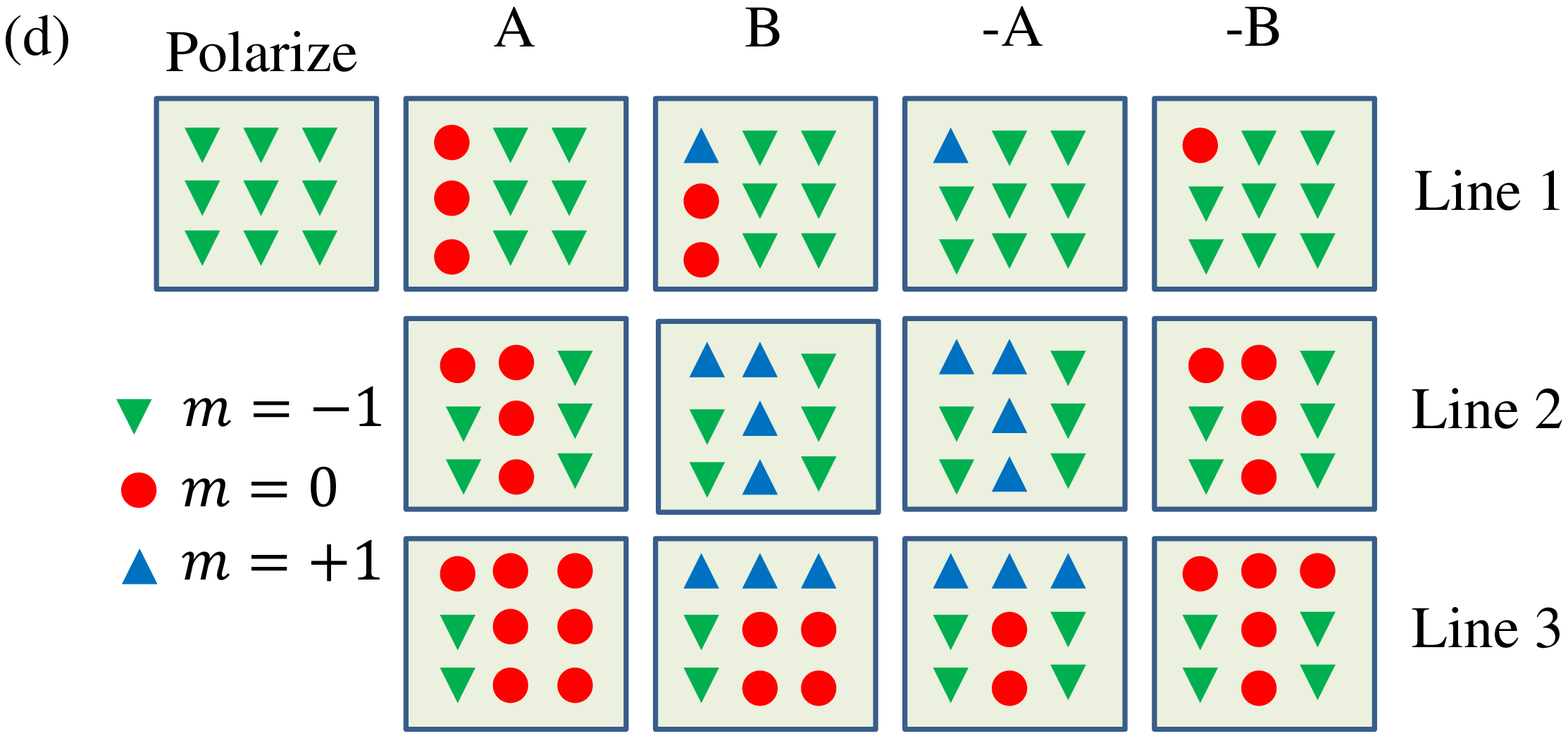}
\caption{(Color online) (a) Schematic setup for magnetic resonance lithography. $B_{0}$ is static magnetic field applied to split the magnetic sublevles $m=\pm 1$ while $B(\nu)$ is microwave applied to couple the transition $m=0$ to $m=\pm 1$; (b) Energy levels for MRL; (c) The pulse sequence to write an arbitrary 2D binary pattern. (d) An example showing how to write a letter ``T" without crosstalk between lines. }
\end{figure}

To see how MRL can achieve high resolution, let us first consider how to write a narrow line in a spin pattern. If we apply a gradient magnetic field, the Zeeman shift of the $m=+/-1$ sublevels is a function of position. A narrowband microwave $\pi$ pulse is  applied to excite the spins from $m=0$ to one of the $m=+/-1$ sublevels. Only spins with the proper Zeeman shift are resonant with the microwave and the others are not appreciably excited. This is similar to the ``resonant z-slice" in MRI \cite{Haacke}. The absolute resolution limit of this technique is given by \cite{McFarland}
\begin{equation}
\Delta x=\frac{\Gamma}{d\Delta_{ZM}/dx}
\end{equation}
where $\Gamma$ is the spin transition linewidth and $d\Delta_{ZM}/dx$ is the spatial gradient of Zeeman shift. The width of the line is not limited by the wavelength but the spin transition linewidth and the gradient of the magnetic field.  In principle, we can achieve arbitrary resolution by increasing the gradient of the magnetic field. For example, if the dephasing time is about 10$\mu$s \cite{Ofori-Okai} and the gradient of the magnetic field is $100Gauss/\mu m$, the resolution can be a few nanometers.

We can also address many spatial locations in parallel as in conventional MRI where the narrowband microwave is replaced by a pulse with more complex spectral contents \cite{Hemmer_spie}. For example, if we want to print an arbitrary 1D pattern characterized by $f(x)$ whose value is normalized into 0 to 1 range, we should have excitation probability given by \cite{Scully}
\begin{equation} 
f(x)=\frac{1}{2}\{1-\cos[\Omega_{R}(x)\tau_{p}]\}
\end{equation}
where $\tau_{p}$ is the pulse duration and $\Omega_{R}(x)=\mu_{ab}{\cal E}(x)/\hbar$ is the resonant part of the Rabi frequency seen by atoms at the position $x$. Here, ${\cal E}(x)$ is the amplitude of the microwave frequency component which is resonant to the transition frequency $\omega(x)$ at position $x$, $\mu_{ab}$ is the transition dipole moment between the two relevant energy levels, and $\hbar$ is the Plank constant.  The transition frequency at position $x$ is given by $\omega(x)=\omega_{0}+g\mu_B B(x)/\hbar$ where $\omega_{0}$ is the zero field splitting, $g\mu_B=2.8MHz/Gauss$, and $B(x)$ is the magnetic field at position $x$. The pulse in the time domain is the Fourier transformation of the required spectrum which in turn is related to position because of the magnetic gradient. 

2D patterns can also be printed in this setup where all three energy sublevels are involved and both X and Y magnetic gradients should be applied. The proposed pulse sequence is shown in Fig. 1(c). For example, we want to write a letter ``T" (Fig. 1(d)). First, all the spins are polarized to the state $|-1\rangle$ (green down triangles). This is the initial state. To write column 1, the X-gradient magnetic field is turned on and a narrowband microwave $\pi$ pulse (A), whose frequency is tuned to be resonant with spins in column 1, rotates spins in this column from $|-1\rangle$ to $|0\rangle$ state.  Then the X-gradient is turned off and the Y-gradient is turned on and another microwave pulse(B) with the desired spectrum rotates the spin in the first row from $|0\rangle$ to $|+1\rangle$ (blue up triangles) state. To write another column without crosstalk, we need to turn on the negative X-gradient magnetic field again (and turn off the Y-gradient) and then apply the negative of pulse A to rotate unwanted spins (red circle dots) in column 1 back to the state $|-1\rangle$ (green down triangles). Finally, we again turn on the negative Y-gradient (with X-gradient off) and the negative of pulse B is applied to rotate the spins in state $|+1\rangle$ back to state $|0\rangle$. A 1D pattern is then formed in the column 1 on the $|0\rangle$ state. After that we can move to the next column and repeat the above procedures. As seen in the figure, the desired 2D pattern ``T" can then be encoded in the $|0\rangle$ state. Finally, an optical readout of the spin population can excite the spins in state $m=0$ into a fluorescing excited state. If the photoresist is within the near-field region of the spins, instead of fluorescing to free space, most of the optical energy of the fluorescing excited state is absorbed by the photoresist through F\"{o}ster resonance energy transfer (FRET) \cite{Clegg, Tisler}.

The FRET efficiency is given by $e=[1+(R/R_0)^{6}]^{-1}$, where $R_0$ is the critical FRET radius and $R$ is the distance between the emitter and the receiver. To capture the high resolution features of the patteren, the photoresist needs to be within the FRET radius from the paramagnetic system. This could be achieved in several ways depending on the nature of the paramagnetic system. For example, for paramagnetic defects in crystals, this could be achieved by shallow implantation or engineered growth techniques. A more practical approach, however, would be to coat the substrate surface with a resin that contains paramagnetic molecules or nanoparticles with paramagnetic defects.

\begin{figure}
\includegraphics[width=0.8\columnwidth]{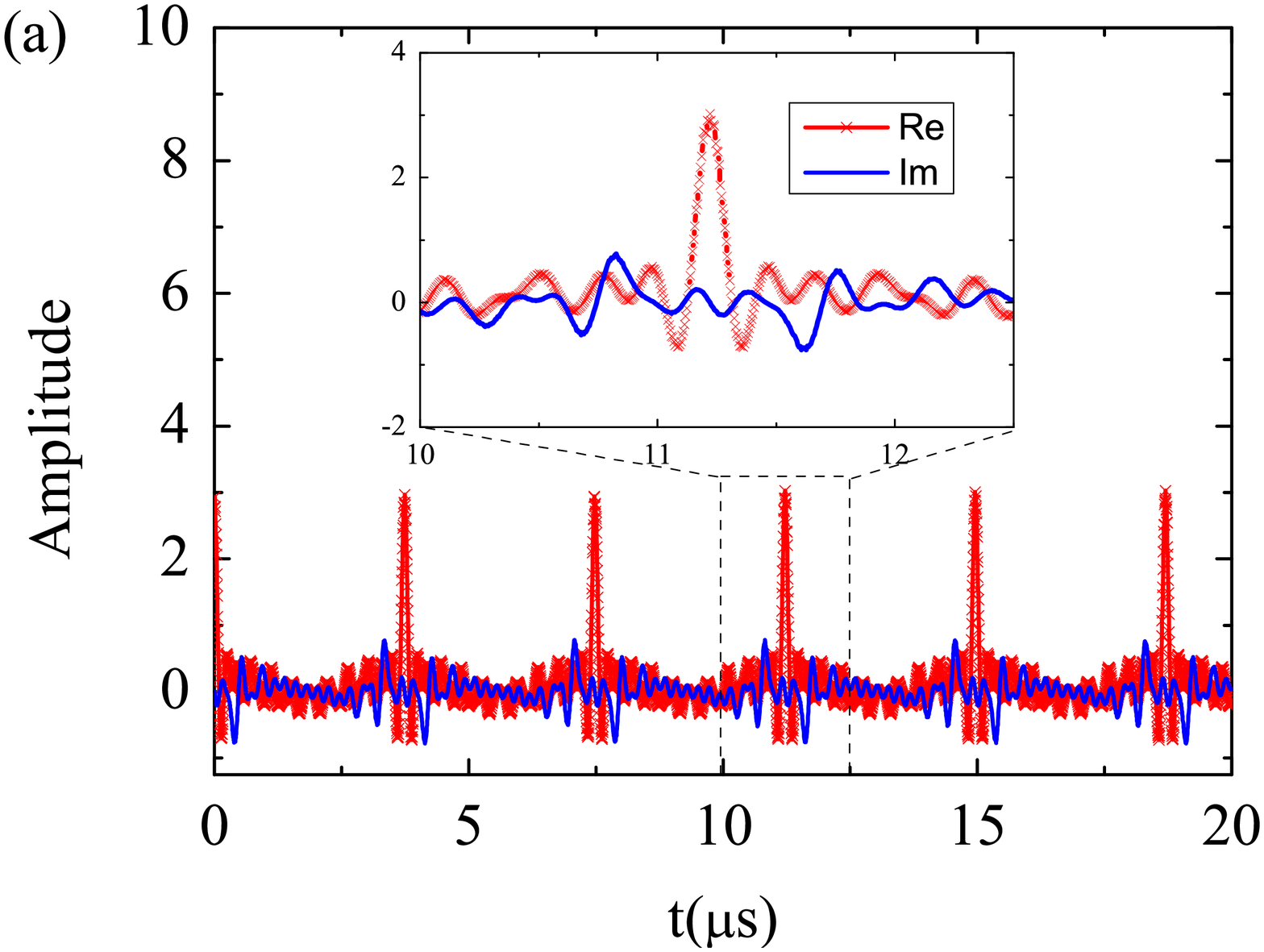}
\includegraphics[width=0.8\columnwidth]{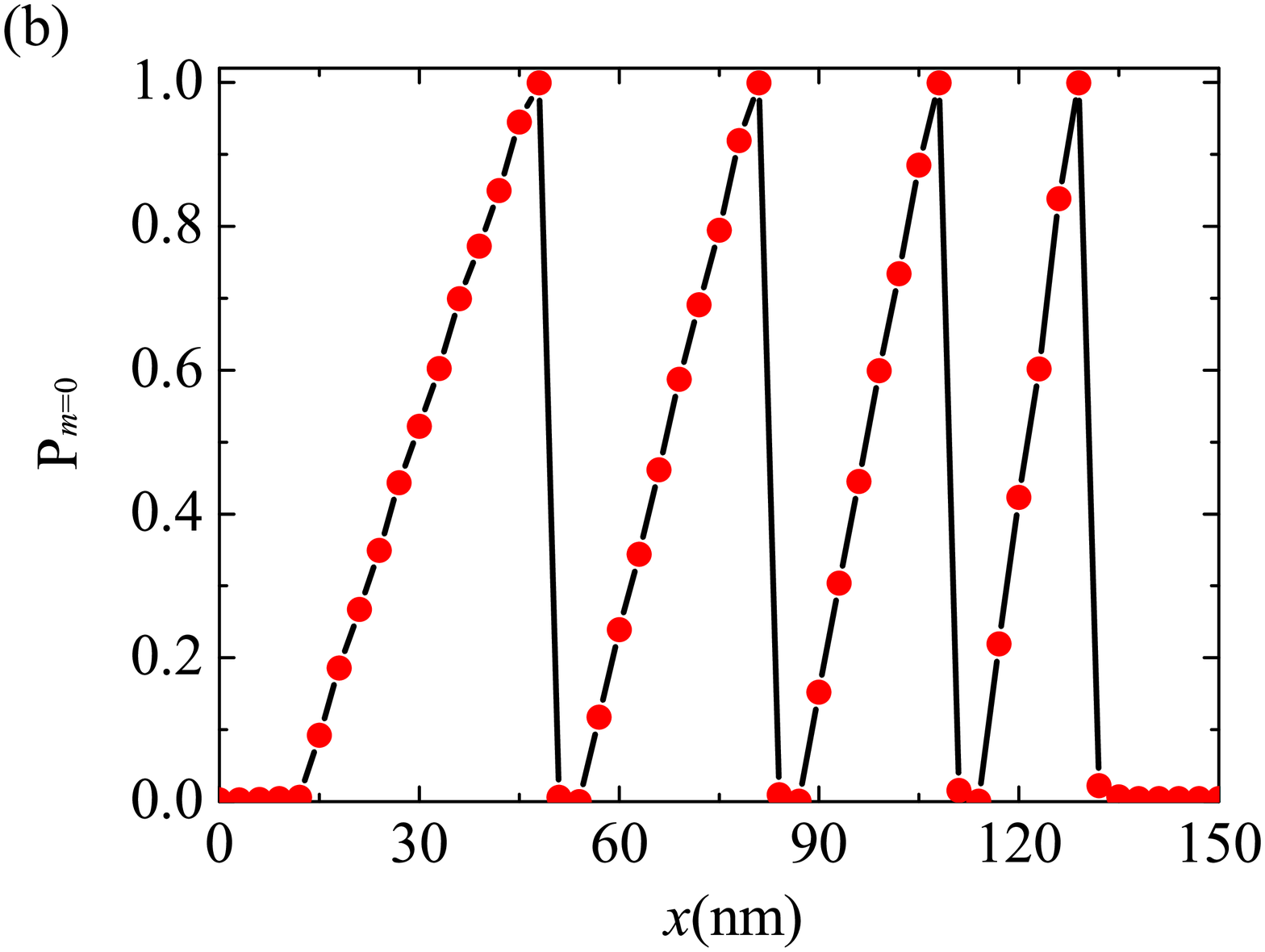}
\includegraphics[width=0.8\columnwidth]{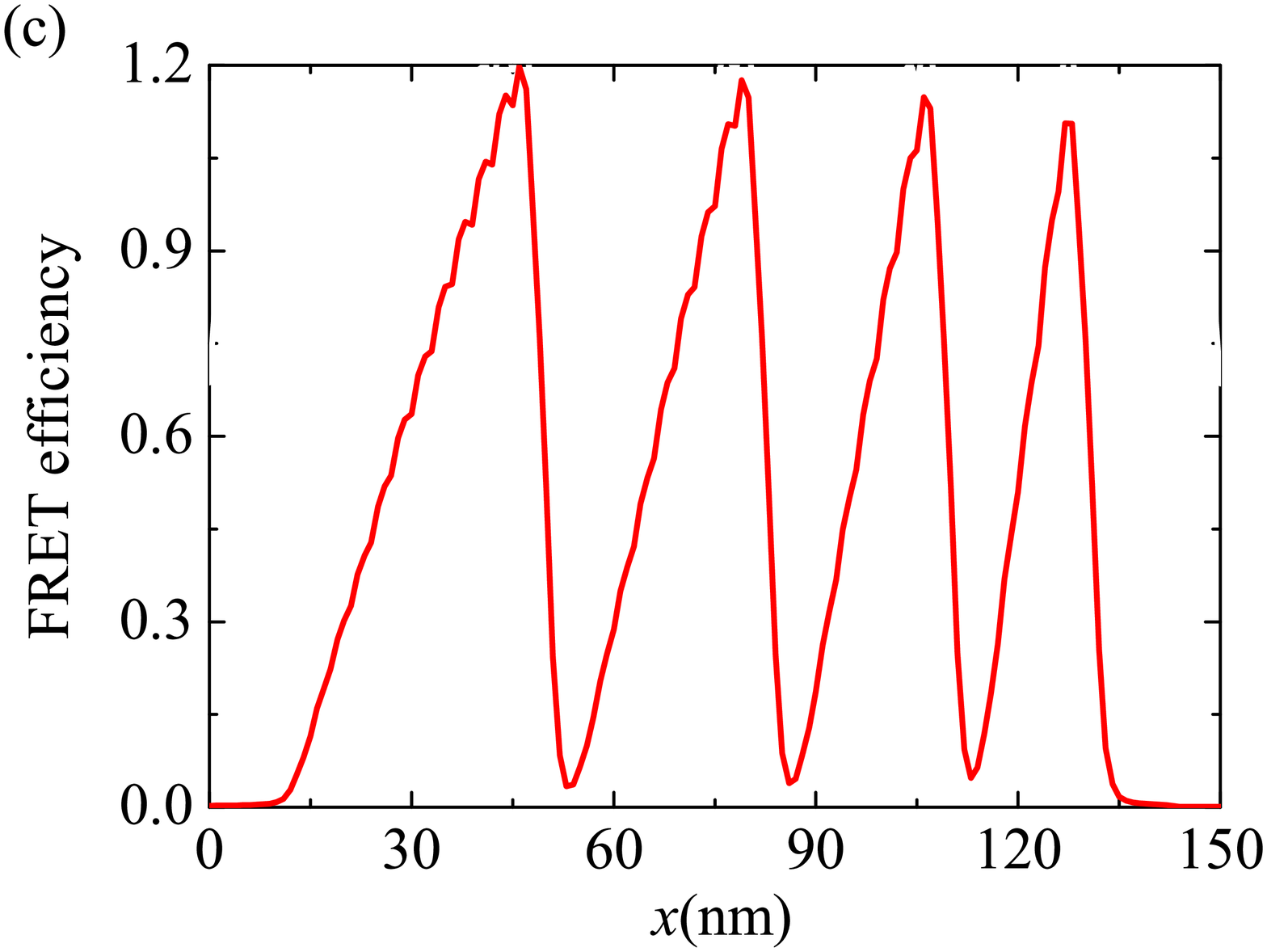}
\caption{(Color online) (a) Complex RF pulse used to write a chirped grating. Red line with cross symbols is the real part while blue line without symbols is the imaginary part. (b) The spatial distribution of the probability that the spins are prepared in the bright state. (c) The FRET efficiency on the substrate surface when the readout contrast is $100\%$. }
\end{figure}

\section{numerical simulation}

In the following, we apply a numerical simulation to write a 1D and a 2D pattern onto the $|0\rangle$ state using MRL. The dynamics of the system is govern by the master equation
\begin{equation}
\dot{\rho }(x,y,t) =-\frac{i}{\hbar}[H(x,y,t),\rho(x,y,t) ]-{\cal L}_{\rho }
\end{equation}
where $\rho$ is $3\times 3$ density matrix for the three energy sub-levels, $H$ is the Hamiltonian which is given by
\[ \frac{H(x,y,t)}{\hbar} = \left[ \begin{array}{ccc}
\Delta_1(x) & \Omega_{1}(t) & 0 \\
\Omega_{1}(t) & 0 & \Omega_{2}(t) \\
0 & \Omega_{2}(t)& \Delta_2({\bf y}) \end{array} \right],\] 
where $\Omega_{1}(t)$ is the Rabi frequency of the $|-1\rangle\rightarrow |0\rangle$ transition and  $\Delta_1(x)$ is the corresponding detuning. $\Omega_{2}(t)$ is the Rabi frequency of the $|0\rangle\rightarrow |+1\rangle$ transition and  $\Delta_2(y)$ is the corresponding detuning. 
The last term ${\cal L}_{\rho }=(\rho _{0,\pm 1}+\rho_{\pm 1,0})/T_2$ is the dephasing of the system. The whole process needs to be done before the population decay time $T_{1}$ which can be of the order of milliseconds. Here we will neglect population decay but keep the dephasing process. In the numerical simulation, we apply the Fourth-Order Runge-Kutta method to solve the time evolution of Eq. (3) with Rabi frequencies and detunings vary as the pulse sequence shown in Fig. 1(c).

For the 1D pattern example, we write a chirped grating within a 150nm field of view using numerical simulation. Here, only an X-gradient and the $|-1\rangle\rightarrow |0\rangle$ transition are involved. The spins are initially prepared in $|-1\rangle$ state. We assume the wire diameter is 1$\mu$m and the distance between the centers of the two wires is 2$\mu m$. The currents in both wires are 50mA and in the same direction. The currents generate a magnetic gradient of about 200Gauss/$\mu$m in the X-direction. The RF magnetic field envelope of the microwave excitation pulse for the $|-1\rangle\rightarrow |0\rangle$ transition in the time domain is shown in Fig. 2(a). The total time duration is about 20$\mu$s. This desired pulse could be generated with a commerical arbitrary waveform generator (AWG) \cite{AWG}. Using an AWG also allows any non-uniformities in magnetic bias or gradient fields to be compensated by applying the appropriate inverse non-linearity to the AWG instruction set. In the numerical simulation, we assume the Rabi frequency has $10\%$ amplitude fluctuation. The paramagnetic material is assumed to be $5$nm deep. The probability that the spins are in the $m=0$ state is shown in Fig. 2(b) where we can see that the desired pattern is encoded in the $m=0$ bright state. The pattern of the internal state can then be transfered to the photoresist by the FRET. The FRET efficiency on the substrate surface is shown in Fig. 2(c) where $R_{0}$ is assumed to  be $5nm$. We can see that a chirped grating can be generated with very high contrast. 

\begin{figure}
\includegraphics[width=0.8\columnwidth]{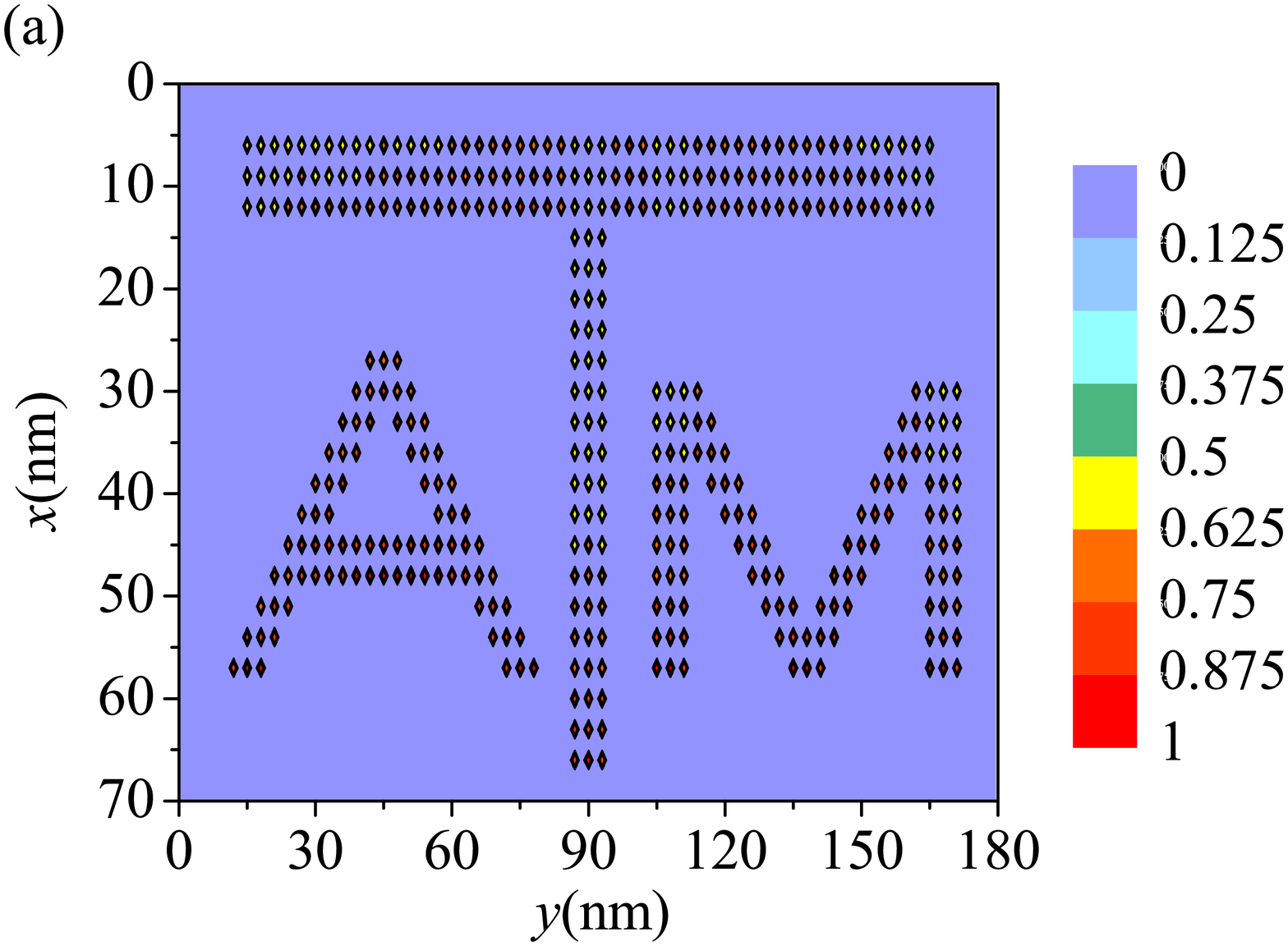}
\includegraphics[width=0.8\columnwidth]{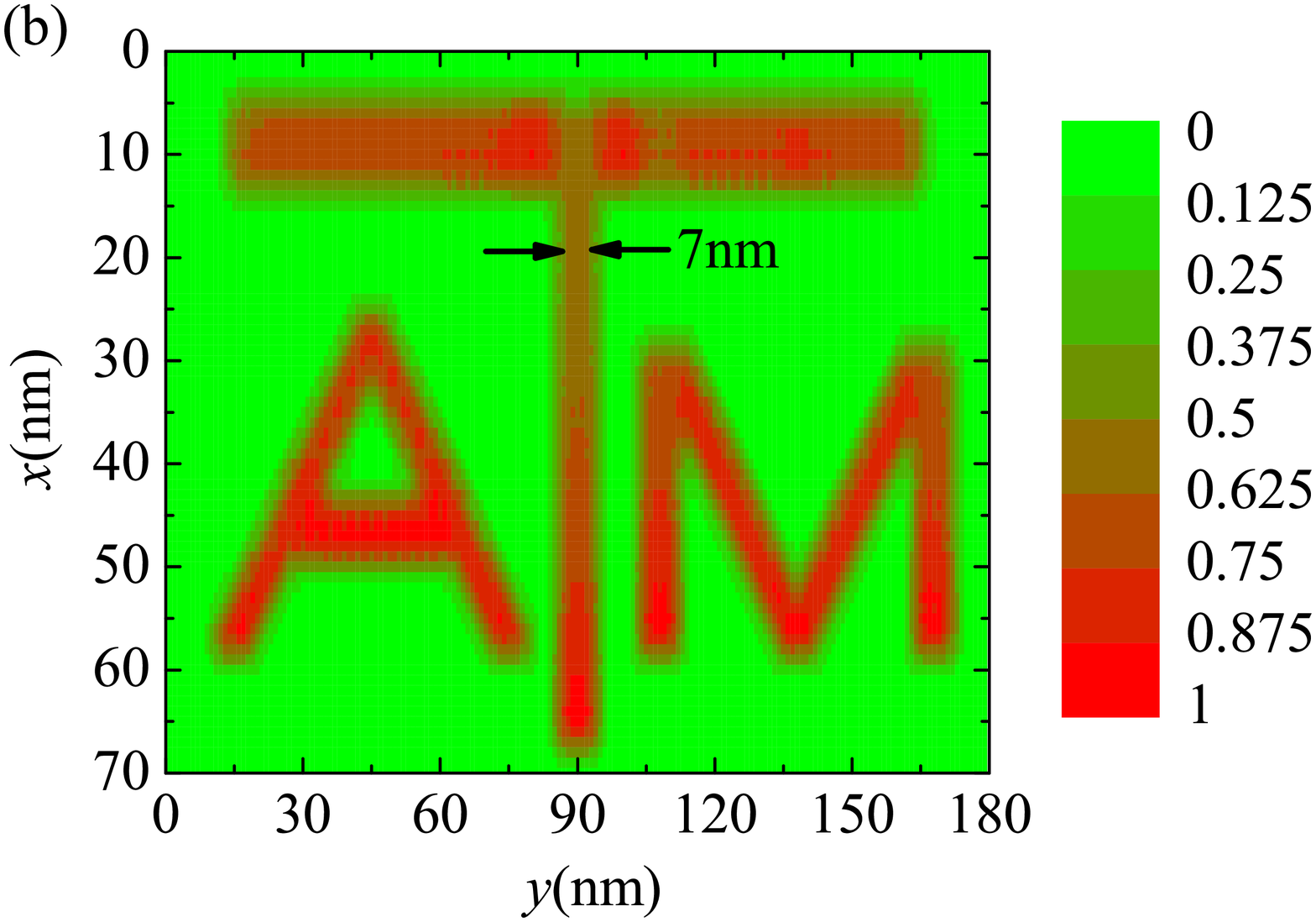}
\caption{(Color online) Numerical simulation to write the Texas A\&M University logo into a two-dimensional area using MRL. (a) The spatial distribution of the probability that the spins are prepared in the bright state. (b) The FRET efficiency on the substrate surface when the readout contrast is $100\%$. }
\end{figure}

For a 2D pattern numerical example, we write the Texas A\&M University logo (``ATM") into a 72nm$\times$180nm region. This area is divided into 24 slices in the $x$ direction and 60 slices in the $y$ direction. The separation of the spins is chosen to be 3nm on a square grid. Each line consists three spins. The pulse sequence is shown in Fig. 1(d) where both X-gradient and Y-gradient magnetic fields are involved. In the numerical simulation, the gradients in both directions are generated by 100mA currents using the same wire separation as in the 1D example. The narrow band microwave $\pi$ pulse used to drive transitions $|-1\rangle\rightleftharpoons |0\rangle$ is assumed to have a Gaussian time envelope. The time duration of the entire pulse A which selects a line is about 5.6$\mu$s where the Gaussian envelope portion has a  width of about 1.5$\mu$s. The time duration of the complex pulse B is also about 5.6$\mu$s which is discretized into 7000 time steps. The total time duration to write one line is about 22.4$\mu$s and it is about 500$\mu$s for the whole process. The dephasing time is assumed to be about 20$\mu$s \cite{Ohno}. Again, the amplitudes of the pulses have a fluctuation of about $10\%$.  The probability that the spins are in the bright state ($m=0$) is shown in Fig. 3(a) where we can see that the desired pattern is successfully encoded into the internal state of the system. Finally, an optical readout can transfer the pattern of the internal states to the photoresist. Assuming that the paramagnetic systems are about 5nm deep and the FRET radius $R_{0}$ is also $5nm$, the relative FRET efficiency at the surface is shown in Fig. 3(b) where we see that the desired pattern can be formed. The full width at half maximum of the line in the middle is about $7$nm.

\section{Possible materials for MRL}

As is mentioned in Sec. II, MRL requires a solid system with high spin transition quality factors and high contrast optical readout of the spin state. There are a number of possible systems which can satisfy these requirements. One promising candidate is the negativley charged Nitrogen Vacancy (NV) color center in diamond. The NV exhibits an optically detected magnetic resonance (ODMR) spin system with long coherence times, even at room temperature \cite{Gaebel,Childress,Wrachtrup,Balasubramanian}. Coherence times larger than a millisecond have been experimentally achieved for NVs leading to their frequent use as nanoscale magnetic field sensors \cite{Balasubramanian}. Such properties could give MRL resolutions as high as a few nanometers. Unfortunately at room temperature the optical readout of the NV spin state has less than $30\%$ contrast. Nonetheless the inherent nonlinearity of resist exposure may allow us to produce high contrast patterns even with this relatively low optical contrast.

One alternative candidate is the silicon vacancy (SiV) center in diamond. The SiV has an optical readout contrast near $100\%$ at low temperature since its ground state sublevel transitions can be optically resolved \cite{Muller2014}. Unlike NV and many other materials, this high resolution persists even for ensembles, due to the low inhomogeneous broadening and low spectral diffusion of SiV. Granted that working at low temperature might produce some technical difficulties, nonetheless photoresist exposure has been demonstrated in cryogenic temperatures \cite{LowTempPhotoresist}. Another important advantage of the SiV is that it can have optical Raman transitions connecting the ground state sublevels. Here light shift gradients could also replace magnetic field gradients \cite{Gardner} so that no wires would be needed on the substrate at all. 

Finally we briefly consider other possible materials systems for performing MRL at room temperature. One of these is the ST1 color center exhibiting a high optical readout contrast up to $50\%$ albeit with shorter spin lifetimes  \cite{Lee2013}. Ruby is also of interest as it has shown room temperature optical Raman transitions with readout contrast comparable to NV\cite{Kolesov}. Other alternatives are the rare-earth doped materials. For instance, super-resolution microscopy has been demonstrated using praseodymium-doped yttrium aluminum garnet (Pr:YAG) at room temperature \cite{PrYAG}. The upconversion UV emission of (Pr:YAG) is optimal for photoresist exposure. Hence the prospects are good that a material suitable for room temperature MRL will eventually be identified. 

\section{summary}

In summary, we have proposed an approach for optical lithography which is the inverse of magnetic resonance imaging. The technique uses atomic coherence in an ensemble of spin systems whose final state population can be optically detected. We design a pulse sequence which can write arbitrary 1D and 2D patterns into the spin states. We presented a numerical simulation to prove the fundamental physics of MRL which shows that our method here is capable of producing arbitrary 1D and 2D high-resolution patterns in one of the spin sublevels with high contrast. The spin pattern can then be transfered to the photorsist pattern by FRET. We also discuss several possible materials for our MRL method.

\section{Acknowledgement}
This work is supported by grants from the Qatar National Research Fund (QNRF) under the NPRP project 5-102-1-026 and 
the King Abdulaziz City for Science and Technology (KACST), DARPA QuASAR, and NSF 1202258. We acknowledge helpful discussions with Charles Santori of HP Laboratories.

\end{document}